# Leveraging User Experience and Learning Analytics for Enhanced Student Well-being

Khadija El Aadmi-Laamech[1], Patricia Santos[1], Davinia Hernández-Leo[1]

[1] Universitat Pompeu Fabra, Pl. de la Mercè, 10-12, 08002, Barcelona, Spain
```
{khadija.elaadmi, patricia.santos,
   davinia.hernandez-leo}@upf.edu
```

**Abstract.** This study explores the design and preliminary evaluation of the "Well-being Journey" (WB Journey), a digital tool aimed at enhancing student well-being within educational environments through tailored recommendations for students. The study examines the WB Journey prototype's user experience and its effectiveness in meeting learning analytics goals related to student preferences. To achieve both goals, we employ a mixed-methods approach, combining quantitative data from the User Experience Questionnaire (UEQ) and the Student Expectations of Learning Analytics Questionnaire (SELAQ) with qualitative feedback from a student discussion. Conducted among 25 students from an engineering school in a Spanish University, the study's data collection involved a 120-minute workshop. The findings suggest opportunities for enhancing the prototype, highlighting the importance of aligning similar digital tools with student needs and preferences for a supportive learning environment, which can be achieved by leveraging tools such as UEQ and SELAQ.

## 1    Introduction

The increasing recognition of well-being as a critical component of student success in higher education [1] highlights the necessity for developing evaluative well-being instruments that are both effective and sustainable [2, 3, 4, 5]. Moreover, the integration of such instruments into educational systems should be seamless and supportive, ensuring that they not only diagnose potential well-being issues but also guide interventions and track progress over time [6, 1]. In doing so, educational institutions can move beyond traditional academic metrics (typically used in Learning Analytics) to embrace a more inclusive and emotional view of learning success, one that encompasses the well-being of students as essential to their academic and life achievements [7, 8, 9].

In this context, the principles of the Self-Determination Theory (SDT) offer a robust framework for understanding and enhancing student well-being within educational contexts. SDT [10] posits that fulfilling basic psychological needs for autonomy, competence, and relatedness is essential for fostering intrinsic motivation, well-being, and personal growth. In the realm of education, this theory highlights the importance of creating learning environments that support these needs, thereby promoting students' self-motivation and engagement with their learning environment [11, 12]. By applying SDT as a guiding philosophy for the development and implementation of well-being instruments, educational systems can more effectively address the well-being needs of students [13, 10, 14]. Furthermore, integrating the SDT into the assessment and enhancement of student well-being ensures that educational practices not only aim at academic excellence but also at cultivating environments where students feel empowered, capable, and connected [14, 15]. This approach aligns with the broader objective of education to prepare individuals for fulfilling lives, both within and beyond the classroom walls.

Therefore, this work delves into the design and preliminary evaluation of a digital tool aimed at assessing and promoting student well-being both on an individual as well as class basis: The Well-being Journey (WB Journey). By leveraging the principles of SDT, this research seeks to provide a nuanced understanding of students' psychological needs, through collecting data about their learning experiences and offering tailored recommendations to enhance their well-being within an educational environment. The tool is designed to fulfill a similar set of objectives for students as those associated with the Learning Analytics Dashboards (LADs), which [16] categorize into six primary goals, with the last four ones also intended for teachers: (1) to enhance retention and academic performance, (2) to assist students in understanding their contribution to group work for its improvement, (3) monitor student interactions within digital learning platforms, (4) offer visual representations of learning outcomes alongside class or group comparisons (5) facilitate student self-reflection and awareness regarding their learning processes and (6) encourage reflection on and awareness of their activities. And while the WB Journey does not directly offer features to cover the two first goals, several studies have stated that a positive well-being in the learning experience is linked to contribute not only to personal flourishment but also to learning outcomes [17, 8, 18, 19], both individually and collectively as a class. As for the four latter goals, the WB Journey aims to achieve them through several features: goal 3 → to monitor student interactions within digital learning platforms through digital well-being evaluations (i.e. readily available tools to evaluate the digital aspects of learning); goal 4 → offer visual analytics depicting students' well-being evolution over time regarding their learning experiences; goal 5 → generate recommendations and actionable feedback to facilitate student' self-reflection and well-being improvement; goal 6 → similar to goal 5, the given recommendations will facilitate reflection and awareness of their learning activities and experiences.

We outline the methodological framework used to answer the following research question (RQ) "What are the essential features that should be integrated into a tool aimed at supporting students' well-being in classroom settings, as determined by students' perceived value of functionalities?", focusing on two research goals (RO): The first goal (RO1) lies on evaluating the design of the WB Journey prototype from a user experience (UX) perspective. The second goal (RO2) lies in achieving the aforementioned LAD set of goals. The instruments utilized to achieve each objective are the following, respectively: 1) the User Experience Questionnaire (UEQ) which helps with guiding the WB Journey prototype's design, emphasizing the importance of a user-centered approach in creating digital interventions that resonate with students' lived experiences as well as developing design-informed decision [20, 21, 22, 23];

and 2) The Student Expectations of Learning Analytics Questionnaire (SELAQ), which offers researchers and practitioners a method for assessing students' expectations regarding learning analytics services (e.g LADs) [24]. As [42] discusses, collecting well-being data can enhance support for students and staff by pinpointing stressors, streamlining information access, offering alternatives, and delivering swift feedback. These goals are aligned with the objectives of the LA field [16].

The discussion extends to the implications of such digital tools for educational practice, advocating for the integration of well-being assessments into the broader educational strategy with students' interest, preferences and expectations in mind, as well as from an ethical and data privacy perspective.

## 2      Background

Student well-being has long been researched and recognized as a critical component of student success [1]. There are multiple dimensions to student well-being that have been explored; as per [25], students' well-being is mainly associated with school connectedness, academic efficacy, joy of learning, and educational purpose, with strong associations with other student well-being indicators; [26] define the dimensions of student well-being in four, these being mental well-being, cognitive well-being, social well-being, and physical well-being; [27] argue that student well-being is highly associated with motivation, teacher behavior, and achievement; [28] differentiate between positive (attitudes, enjoyment, self-concept) and negative (worries, physical complaints, social problems) student well-being dimensions.

Naturally, when devising instruments to measure student well-being, these highly depend on the informed dimensions, and each primarily tied to context age, gender and school climate [29]. Through the dimensions and instruments for well-being, one particular well-being theory stands out due to its approach to the understanding of well-being: The SDT. Unlike other well-being theories, the SDT does not rely solely on well-being dimensions, but breaks down well-being to three essential needs, also known as basic psychological needs: competence, autonomy and relatedness [10]. These needs transcend across the different instruments that originated from the SDT, making it quite versatile in the various domains and contexts it is used in, but with the common denominator of the BPNs [10]. This characteristic can help with the evaluation of the different components of the learning environments that the students may interact with. For instance, as per [30] the "SDT explains the factors of intrinsic and extrinsic motivation that support personality development and behavioral self-regulation, improving personal well-being and performance in organizations and society". Furthermore, the SDT has also been proven to promote "interest in learning, valuing education, and confidence in one's own abilities, leading to high-quality learning and personal growth" [31].

The use of the SDT in education has also extended to Technology Enhanced Learning (TEL), proving its integration within technological learning environments to be just as useful and effective as regular learning environments, with recent literature highlighting the SDT's significant role in digital education [32]. Though current literature advocates for the use of technology to enhance learner well-being whilst reducing its negative impacts [33, 34], the integration of the SDT becomes of particular interest due to its usefulness in improving well-being whilst also diagnosing potential adverse impacts of technology. Furthermore, [35] discuss the potential of digital interventions in improving student well-being, but also highlight the need for more research focusing on effectiveness as well as user experience of such interventions. In that regard, this work on the design of the WB Journey not only aims at discussing the development of digital interventions and recommendations with the objective of improving student well-being, but also aims to contribute to the research agenda regarding such well-being tools' effectiveness and user experience.

## 3      The Well-being Journey prototype

The WB Journey prototype is designed to primarily evaluate both the well-being of students individually as well as the well-being of the class as a whole. The idea of the WB Journey was born following the current need of developing evaluative well-being instruments that are both effective and sustainable in educational contexts [2, 3, 4, 5]. It offers both digital and non-digital well-being evaluations, with the intent to grow the repertoire of instruments it provides. These instruments are mainly SDT-based, which means that they focus on BPNs satisfaction. Though the objective of the WB Journey is to evaluate student well-being, it is up to the teacher to choose which evaluative instrument to use from the provided repertoire, based on the environment the teacher wants to evaluate (e.g. the evaluation of student well-being when using a technological tool, the evaluation of well-being of the classroom dynamics, etc.). Based on the results of using the provided instruments, students are then given a set of recommendations that they can apply to their learning experience in order to actively improve their well-being. Furthermore, the WB Journey aims to create a sustainable cycle of well-being improvement, meaning the teacher can monitor the well-being of the class by taking the same test during different moments of their course. This way, students will also be able to visualize their well-being evolution, if they wish so.

### 3.1      WB Journey Interface functionalities and details

The prototype is designed in a way that makes the flow of creating or taking well-being tests as easy and intuitive as possible. The home page contains two types of accesses: 1. Access for teachers/designers: they can access the prototype and create tests for their students to take. 2. The second type of access is directed to students, where they use a code the teacher generates beforehand through the first type of access. To create a test, figure 1a represents type of template selection and customization. Teachers are given the option to customize templates (figure 1b); in case a question does not fit their students' needs and/or context, teachers can opt to remove it. Otherwise, the teacher can select using the regular template (figure 1a).

Once a test is created, students can access the test by entering the code the teacher generated, directly in the homepage (refer to figure 1c). The process of how students take the test is reflected in figure 1d. Once students submit their answers, their well-being score is obtained by calculating the central tendency (mean) of each BPN individually as well as aggregated.

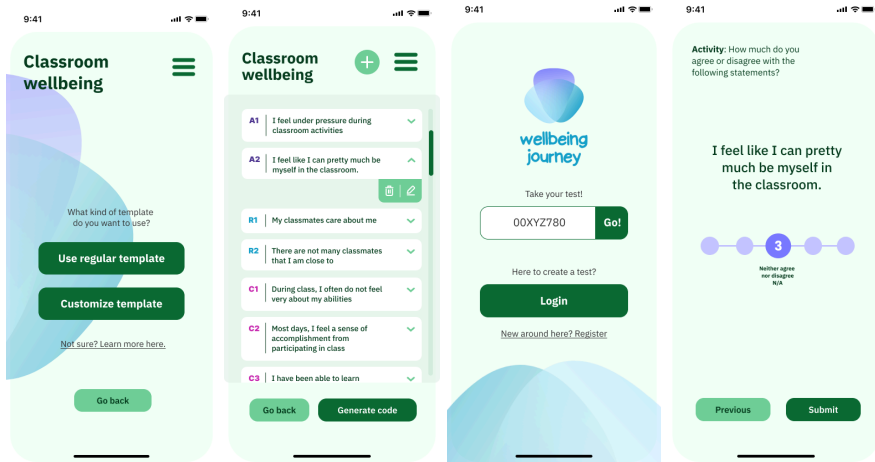

**Fig. 1a.** Template selection, **1b.** Template customization  **1c.** Enter code, **1d.** Question display example.

Afterwards, the system gives out recommendations based on the analysis of each student's well-being score, as well as the class as a whole. Recommendations are generated based on the lowest-scoring (negative well-being) items of each BPN. It is important to mention that the recommendations the system uses are generated by educational experts that studied the SDT-base evaluative instrument beforehand, and created a list of recommendations based on each evaluated item. For instance, the first iteration of recommendations included a total of 23 experts from various subfields of TEL. The figures (2) below are a representation of how the recommendations system works.

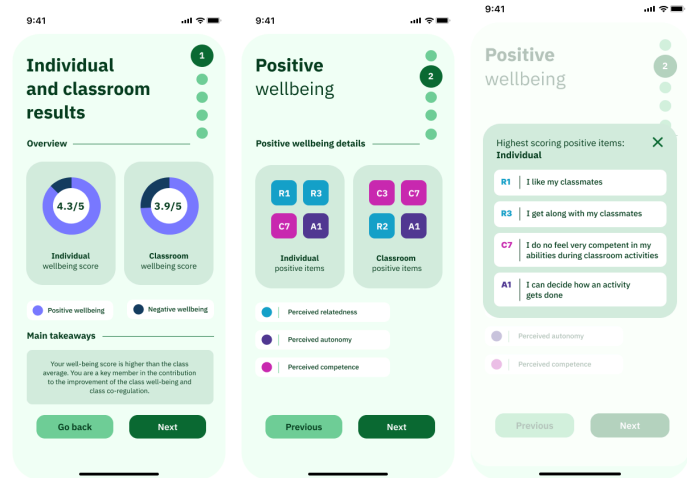

**Fig. 2a**. Main results, **2b.** Positive well-being results, **2c.** Individual positive results

Furthermore, aside from the recommendations (figure 2g) that the system generates, the students have the option to visualize the evolution of their well-being, compared with other classes within the same course (figure 2f). All data that is given after each test can be downloaded in the last step of the results section, in PDF format.

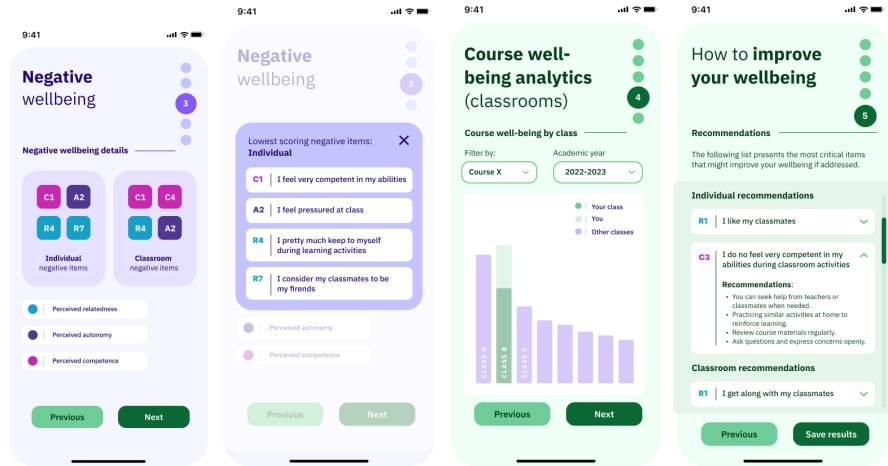

**Fig. 2d.** (negative well-being results), **2e.** (detail of individual negative results), **2f.** (course comparison), **2g.** (recommendations page)

## 4    Evaluation methodology

The primary objective of this study is twofold, centered in the student profile: to evaluate the WB Journey prototype's design from a User Experience perspective and to assess its effectiveness in meeting the set goals similar to those of Learning Analytics Dashboards (LADs), focusing on student well-being enhancement as well as their user needs. We use a mixed methods approach: we quantify the UX and LADs objectives through the User Experience (UEQ) questionnaire and the Student Expectations of Learning Analytics Questionnaire (SELAQ) respectively, accompanied by a qualitative evaluation that aids in collecting suggestions and improvements from students. The data collection was organized as a face to face session of 120 minutes of duration: 1. Introduction to the WB Journey + navigation (30min), 2. UEQ (20 min), 3. SELAQ (20 min), 4. Student discussion (30 min), 5. Logistics and miscellaneous (student arrival, compensation, leaving, etc.) (20 min).

The workshop recruited 25 engineering education student volunteers at a Spanish public University. All students were course peers, which helped the discussion be more fluid.  Students were compensated for their participation and completed a consent form.

### 4.1    Instruments and methods

**1. User Experience Questionnaire (UEQ)** [21]. The purpose of this instrument is to assess the overall user satisfaction with the WB Journey prototype, focusing on usability, aesthetics, and the effectiveness of tailored recommendations. The procedure is the following: participants will be asked to interact with the WB Journey prototype over a specified period. Subsequently, they will complete the UEQ to provide feedback on their experience.

**2. Student Expectations of Learning Analytics Questionnaire (SELAQ)** [24]. The objective of this tool is to evaluate the prototype's effectiveness in meeting the set of learning analytics goals. Students' expectations and experiences related to learning analytics services will be assessed through SELAQ before (ideal expectations) and after (actual expectations) interacting with the WB Journey prototype. This will help determine the prototype's impact on students' engagement and well-being data. The items used from the questionnaire are: Q1. Ask for consent (identifiable data); Q2. Data will be kept securely; Q3. Consent before outsourcing; Q4. Ask for consent (educational data); Q5. Ask for consent (if the data collection purpose is changed); Q6. The LA service will promote student decision making; Q7. The LA service will show learning progress in regard to learning goals; Q8. The LA service will show learning progress across modules/courses, Q9. The teaching staff will be competent enough to interpret the LA service feedback; and Q10. The teaching staff will act on the LA service feedback and provide support.

**3. Qualitative data collection through a student discussion**. In the final part of the workshop, students were prompted to discuss their opinions, ideas and suggestions for the improvement of the WB Journey prototype. These were collected in written format and analyzed through a thematic analysis.

## 5 Results and analysis

### 5.1 User Experience Questionnaire

The UEQ [21] analysis tool provides a myriad of data calculations and visualizations. We selected the main takeaways and interpreted them following the guidelines provided in the analysis tool, completing the RO1 (evaluating the design of the WB Journey prototype from a UX perspective). First, the evaluated items are clustered in the six following qualities (figure 4).

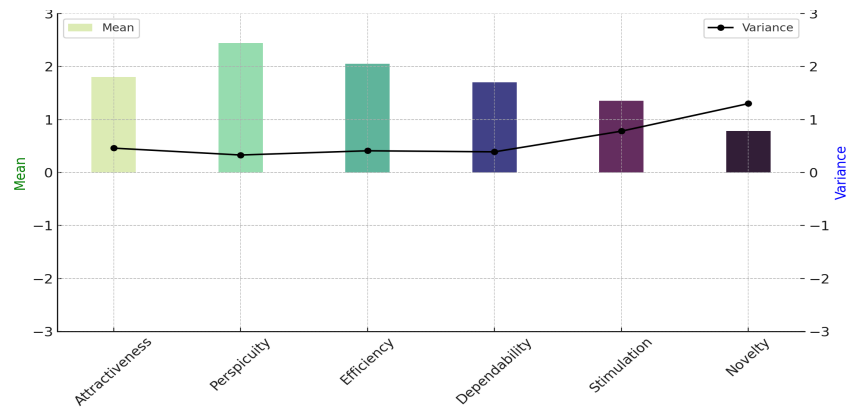

**Fig. 4.** UEQ qualities distribution

The interpretation of the results is as follows. **Attractiveness**: The positive mean (1.80) indicates that students generally find the experience within the WB Journey attractive. The variance (0.46) level shows some variability in perceptions of efficiency, though it remains relatively low. **Perspicuity**: With a high mean (2.44), students find the experience clear and understandable. The low variance (0.33) indicates a strong agreement among students regarding the perspicuity of the experience. **Efficiency**: The efficiency (2.05) of the experience is rated positively, suggesting that students can achieve their goals effectively. The relatively low variance (0.41) suggests that opinions about attractiveness are fairly consistent among students. **Dependability**: Students perceive the experience as dependable (mean = 1.70), but the score indicates there might be room for improvement. The variance (0.39) suggests a general agreement on this aspect, with slight differences in user experiences. **Stimulation**: The experience is somewhat stimulating (mean = 1.35), engaging students to a degree. However, the higher variance (0.78) compared to other scales indicates a broader range of opinions about how stimulating the experience is, suggesting that while some find it very engaging, others do not. **Novelty**: The positive but lower mean (0.78) suggests that the experience has some novel aspects, but it may not be perceived as highly innovative by all students. This is evident in the high variance score (1.30), which indicates significant differences in how students perceive the novelty of the experience, with some finding it less innovative than others.

All in all, the UEQ scale shows that the user experience is generally positive across various aspects, with strengths in understandability, efficiency and attractiveness. However, there are opportunities for improvement, especially in making the experience more stimulating and novel for a broader range of students.

### 5.2   Students Expectations of Learning Analytics Questionnaire

The SELAQ questionnaire is undertaken twice, once before being exposed to the WB Journey (ideal expectations) and after being exposed to the WB Journey (actual expectations). This will help us understand to which extent the WB Journey adjusts to the students' needs and meets the LAD set of goals (RO2).

**Ideal data analysis.** The analysis of the data regarding students' ideal preferences for certain features or policies related to their educational data and the use of a university app shows the following averages (means) and standard deviations for each item (table 3):

**Table 3.** Ideal data scores

|    | Q1   | Q1  | Q3   | Q4   | Q5   | Q6   | Q7   | Q8   | Q9   | Q10  |
|----|------|-----|------|------|------|------|------|------|------|------|
| M  | 4.84 | 5.0 | 4.72 | 4.76 | 4.84 | 4.56 | 4.56 | 4.56 | 4.76 | 4.44 |
| SD | 0.78 | 0.0 | 0.66 | 0.81 | 0.46 | 0.57 | 0.64 | 0.64 | 0.51 | 0.85 |

The highest mean score (5.00) with the lowest standard deviation (0.00) is for ensuring that all educational data is kept securely (Q2). This indicates a unanimous agreement on the importance of data security among respondents. Consent before using identifiable data and before data outsourcing is also highly rated (Q1

mean=4.84 and Q3 mean=4.72, respectively), emphasizing a strong desire for control over personal information.

Promoting student decision making (Q6), and showing progress compared to goals (Q7) all have mean scores of 4.56. This suggests a desire for active and engaging tools within the educational system that support personal and academic growth. However, these areas also show relatively higher variability, indicating differing opinions on their importance.

The competency of the teaching staff in incorporating analytics into their feedback and their obligation to act if analytics indicate a student is at risk both score highly (Q9 mean= 4.76 and Q10 mean= 4.44, respectively), but with notable variability, especially for the obligation to act (Q10 std = 0.85). This highlights the importance of responsive and competent teaching staff while acknowledging some variability in expectations among students.

**Expected data analysis.** The analysis of the students' expectations regarding the university's (when using the WB Journey) handling of student data reveals the following average (mean) scores and standard deviations (table 4). Compared to the ideal scenarios, the actual expectations have lower means across all areas. This suggests an area of overall improvement between what students hope for and what they realistically expect to happen regarding their educational data and app usage. Though as specified by [24], it is common to expect lower means across the board in the expected outcomes compared to ideal expectations.

Table 4. Ideal data scores

|    | Q1   | Q1   | Q3   | Q4   | Q5   | Q6   | Q7   | Q8   | Q9   | Q10  |
|----|------|------|------|------|------|------|------|------|------|------|
| M  | 4.36 | 4.24 | 3.92 | 3.96 | 3.52 | 3.16 | 3.64 | 3.40 | 2.60 | 2.12 |
| SD | 0.69 | 0.95 | 1.23 | 1.08 | 1.36 | 1.01 | 1.02 | 1.02 | 0.98 | 1.11 |

Notably, there's a significant drop in expectations for the competency of teaching staff in using analytics (Q9 mean, from 4.76 to 2.60), and the obligation of teaching staff to act on analytics (Q10 mean, from 4.44 to 2.12). These areas represent the biggest disparities between ideals and expectations, indicating the need for more effective and proactive strategies that can be used by the educational staff in leveraging analytics to support student success.

As for the expectation scores for data security and consent, even though they are lower than the ideal scores, they remain relatively high compared to other areas. This highlights a continued emphasis on privacy and data protection among students.

The standard deviations are generally higher in the expectations data, especially for consent before outsourcing data (Q3 std= 1.23), further consent for different data use (Q5 std= 1.36) and the teaching staff competence to act on the LA service feedback and provide support (Q10 std=1.11). This increased variability suggests that students are more uncertain about what will actually happen in these areas, reflecting a diverse range of expectations.

### 5.3     Thematic analysis

Based on the responses provided to the question about improving the WB Journey, several key themes and open suggestions emerge that focus on enhancing the user experience as well as enhancing and repurposing the obtained recommendations, helping us better achieve the LAD goals. These suggestions were categorized into the following key thematic areas, which do not include simple answers (e.g."I like the app as it is now"), themes that were already addressed by the SELAQ questionnaire (e.g. "consent and privacy") and non-responding students:

**Academic support**: **i) Peer support and advice** (28% of respondents): Some responses advocate for including advice from experienced students (aside from experts), offering peer-to-peer guidance based on personal experiences. Peer support has long been investigated as a contributing factor to student well-being [36, 37, 38]. This helps achieve LAD goals 5 and 6. **ii) Course-specific recommendations** (24% of respondents): There's a clear interest in understanding how specific subjects affect well-being, including the perceived difficulty and stress levels associated with them. This could help in managing expectations and preparation for courses. This helps achieve LAD goals 5 and 6.

**Personalization and tracking**: **i) User profiles and tracking well-being** (28% of respondents): There's a strong desire for personalized tracking of well-being across different time frames (e.g., quarterly, annually) and academic scenarios (e.g., different subjects, class types). This includes comparing one's progress over time and against various benchmarks, suggesting a more personalized and longitudinal approach to monitoring well-being to the one currently provided in the WB Journey. This aligns with recent research that effectively utilized a longitudinal approach to track both well-being and academic performance [6, 1]. This helps achieve LAD goals 3 and 4. **ii) Customizable reminders** (28% of respondents): Students proposed implementing active reminders to help them focus on their well-being and improve self-awareness, as a more engaging option than just providing downloadable resources with their recommendations. [39] address this sense of self-awareness as a positive means to improve student well-being in higher education. This helps achieve LAD goals 5 and 6.

**Educational insights**: **i) Implement comparative features** (32% of respondents): Integrate features that allow students to compare their well-being with peers, including those in similar academic situations or taking the same subjects. This can foster a sense of community and shared understanding among students [40]. This helps achieve LAD goals 4, 5 and 6. **ii) Global and group data access** (28% of respondents): Access to broader data sets, including the well-being scores of different groups, subjects, and academic years, is desired. This wider perspective could aid in setting realistic expectations and understanding general trends in academic well-being. This helps achieve LAD goals 4, 5 and 6. **iii) Recommendations for teachers** (20% of respondents): Respondents seek recommendations targeted at teachers to improve classroom experiences, suggesting a dual focus on both student and teacher well-being and performance. This suggestion is quite critical, since as mentioned in the literature, teacher well-being can be a main contributor to student well-being [27], therefore important to address. This helps achieve LAD goals 4, 5 and 6.

**Technological enhancements (12% of respondents):** Respondents suggest implementing features for students to provide feedback on classes, and for sharing personal improvement recommendations, creating a more interactive and reflective tool. This helps achieve LAD goals 3, 4, 5 and 6.

Furthermore, students were also asked whether they would likely rethink any feature from the WB Journey. Their answers are summarized in the following points:

**Content and terminology adjustments (24% of respondents):** A recurrent theme is the suggestion to rethink sections that currently carry potentially demotivating or negative connotations, such as "Negative Well-being." Suggestions include using terms that focus on improvement or strengths (i.e. from "positive well-being" to "my strengths"), aiming to foster a more positive and constructive mindset.

**Data privacy and anonymity (12% of respondents):** Feedback indicates a preference for making group statistics an optional view rather than a default feature, to avoid unnecessary comparisons that could impact students' well-being negatively.

**Accessibility of results (28% of respondents):** There's a call for making results more accessible and user-friendly, such as by saving them directly in the prototype or sending them to students' emails instead of providing them in PDF format, which is seen as less convenient for frequent access.

## 6      Discussion

The quantitative data from 25 respondents, detailing ideal and actual expectations, coupled with qualitative insights from the thematic analysis, uncover interesting correlations. These findings enrich our understanding of student preferences and concerns about the use of well-being data and analytics implementation in education, particularly within the WB Journey.

Both the quantitative data and the thematic analysis emphasize the importance of data security and obtaining consent before using personal data. The high scores for security and consent in ideal and actual expectations align with the qualitative feedback, highlighting a need for privacy and control over personal information.

As for specific feature improvement, the thematic feedback from students further elaborates on the qualitative aspects of their experiences, advocating for a more personalized, engaging, and comprehensive approach to well-being support. Students express a desire for features such as peer support, personalized tracking, customizable reminders, and educational insights that allow for comparisons with peers in different dimensions (courses, academic years, and so on). These suggestions not only highlight a pathway towards creating a more nuanced and supportive tool that can address the diverse needs and preferences of its users.

The thematic analysis also highlighted a need for teaching staff skilled in integrating analytics into personalized feedback and support. Quantitative data aligns with this, revealing high ideals but notably lower expectations regarding staff competency and their duty to utilize analytics, pinpointing an improvement area for the WB Journey. Addressing this by equipping teachers and staff with essential tools and analytics for timely action could enhance student well-being outcomes.

We further develop these implications in the following Conclusion section.

## 7      Conclusion

This work presented the preliminary design of the WB Journey, a tool aimed at facilitating and supporting student well-being through the application of SDT-based instruments and recommendations. We address the RQ, "What are the essential features that should be integrated into a tool aimed at supporting students' well-being in classroom settings, as determined by students' perceived value of functionalities?". By leveraging both UEQ and SELAQ, as well as a thematic analysis, we identified key areas where the WB Journey can evolve to meet the nuanced needs of its users, and which go beyond simply improving the UX (RO1) and achieving the LAD goals (RO2). Specifically, the integration of well-being targeted Learning Analytics within the WB Journey (or systems that want to integrate well-being elements within their evaluations) presents a promising avenue to enhance the educational experience of students, provided that it is executed with careful consideration of the most critical takeaways for design and strategy:

**Supporting an enhancement in teaching staff competence, responsiveness and well-being** (RO1, RO2): There is a clear demand for improving the competency of teaching staff in utilizing LA to offer meaningful support and feedback. Providing embedded training and resources within the WB Journey can empower educators to more effectively leverage analytics, aligning with student expectations for personalized and proactive support. Moreover, there is also a call for well-being recommendations targeted not only at students but also at teachers, to foster an environment that supports the well-being of both parties.

**Personalization and tracking** (RO1, RO2): Feedback from the thematic analysis revealed a strong desire for more personalized approaches to well-being tracking and support. Incorporating user profiles, customizable notifications, a longitudinal approach to well-being data tracking as well as goal-setting features can make the WB Journey more tailored to individual student needs.

**Community and connectivity** (RO1, RO2): The feedback emphasizes the significance of building a sense of community and connectivity among students. By facilitating peer advice and sharing of experiences, along with providing a platform for teacher-student interaction focused on well-being, the tool can strengthen the educational ecosystem.

**Data privacy and consent** (RO2): Work on providing clearer guidelines on how data is handled by the WB Journey, addressing students' concerns about their data privacy as well as their control over their personal data.

These takeaways connect with the wider scope of research on well-being in education, recognizing that learning outcomes are deeply intertwined with students' psychological and emotional well-being, and offering several key benefits to the TEL field:

**Enhanced learning outcomes**: Research consistently shows that students' well-being is closely linked to their learning efficiency, motivation, and overall academic performance, [28]. By prioritizing well-being, the TEL community can foster environments that not only support academic success but also contribute to the development of well-rounded individuals.

**Personalized learning**: The WB Journey's emphasis on individual and class-wide well-being assessments aligns with the broader TEL movement towards personalized learning [41]. Understanding and addressing the unique well-being needs of each student can lead to more effective and tailored educational experiences.

**Promoting a supportive educational environment**: By enabling teachers to monitor and address the well-being of their students actively, tools like the WB Journey contribute to creating a more supportive and empathetic educational ecosystem [40]. This is crucial for nurturing positive student-teacher relationships and building a sense of community within classrooms.

**Sustainable well-being practices**: The cyclical model of well-being assessment and improvement proposed by the WB Journey, alongside the student feedback and preferences, introduces a sustainable approach to well-being in education [40]. It emphasizes ongoing evaluation and adaptation, which is essential for meeting the evolving needs of students during their educational journey.

These insights call for critical future work, including iterating the WB Journey. The iterations range from integrating the feedback proposed by students, to evaluating the tool from a teacher perspective, especially in understanding how well equipped the WB Journey is in helping teachers in providing student support from the insights the tool generates. Improving these aspects can enhance the support the tool provides as well as the overall use and integration of the WB Journey by both students and teachers in educational settings.

**Acknowledgments.** A third level heading in 9-point font size at the end of the paper is used for general acknowledgments, for example: This study was funded by X (grant number Y).

**Disclosure of Interests.** The authors report no conflict of interests.